\documentclass[conference]{IEEEtran}
\IEEEoverridecommandlockouts

\usepackage{cite}
\usepackage{amsmath,amssymb,amsfonts}
\usepackage{graphicx}
\usepackage{textcomp}
\usepackage{xcolor}
\usepackage{xspace}
\usepackage{url}
\usepackage{booktabs}
\usepackage[T1]{fontenc}
\usepackage{balance}
\usepackage{dblfloatfix}

\newcommand{\stableshots}{\textsc{StableShots}\xspace}
\newcommand{\tvd}{\mathrm{TVD}}

\begin{document}

\title{StableShots: Online Shot Stopping for Quantum Circuit Execution}

\author{
\IEEEauthorblockN{Giuseppe Bisicchia}
\IEEEauthorblockA{\textit{University of Pisa, Italy}\\
giuseppe.bisicchia@phd.unipi.it}
\and
\IEEEauthorblockN{Alessandro Bocci}
\IEEEauthorblockA{\textit{University of Pisa, Italy}\\
alessandro.bocci@unipi.it}
\and
\IEEEauthorblockN{Ernesto Pimentel}
\IEEEauthorblockA{\textit{University of Malaga, Spain}\\
epimentel@uma.es}
\and
\IEEEauthorblockN{Antonio Brogi}
\IEEEauthorblockA{\textit{University of Pisa, Italy}\\
antonio.brogi@unipi.it}
}

\maketitle

\begin{abstract}
Quantum circuit execution estimates output distributions by repeated measurements, yet developers commonly choose a fixed shot budget before execution. This static choice is brittle: low budgets can under-sample the distribution, while high budgets waste measurements. In this paper, we present \stableshots, a black-box online stopping rule for static quantum circuits. The method executes a fixed circuit in small batches, monitors the total-variation distance between cumulative empirical distributions, and stops after repeated evidence of local stability. We evaluate \stableshots on 180 QSimBench traces spanning six circuit families, six sizes from 4 to 14 qubits, and five noisy IBM simulated backends. With validation-only calibration and 100 repeated backend-holdout splits, the selected configuration reaches TVD $\leq0.05$ on all held-out test evaluations with median 7,650 shots, whereas fixed-shot baselines either fail more often or spend substantially more shots. 
\end{abstract}

\begin{IEEEkeywords}
Quantum Software Engineering, Quantum Circuit Execution, Shot Optimization, Empirical Evaluation
\end{IEEEkeywords}

\section{Introduction}
Quantum programs are executed by repeatedly evaluating circuits and measuring their outputs. Each shot samples the distribution induced by the circuit and the backend~\cite{nielsen2010quantum}; consequently, the shot count controls cost, queue usage, and the fidelity of the empirical distribution returned to the developer. Current execution workflows commonly expose shots as a fixed scalar budget, such as 1,000, 5,000, or 10,000 shots. This interface is simple, but it hides a software-engineering decision: the useful budget depends on circuit family, size, backend noise, and output sparsity. A single fixed value can therefore be excessive for one execution and insufficient for another.

This problem is aligned with the concerns of quantum software engineering (QSE). It is not only a statistical question about sample complexity, but also an execution-control question about how quantum software should be run in a systematic, measurable, and auditable way~\cite{bisicchia2024quantum}. We focus on \emph{static quantum circuits}: the circuit structure and parameters remain fixed while shots are collected. Workloads whose target distribution changes by design, such as optimizer-driven variational loops, are outside this scope. For a static circuit, however, the backend-induced distribution is fixed but unknown, so it is meaningful to ask whether the cumulative empirical distribution is still changing as new batches arrive.

We propose \stableshots, a lightweight wrapper around any execution interface that returns batch-level counts. The method requires no circuit inspection, backend calibration data, or noise model. It accumulates counts, compares the current empirical distribution with an earlier cumulative distribution, and stops only after repeated evidence that the marginal effect of new batches is small. The decision is traceable from the observed counts, which makes the policy suitable for QSE settings where reproducibility and execution provenance matter.

This paper concentrates on one research question: \emph{how effectively does \stableshots reach a target distributional-fidelity regime compared with fixed-shot baselines?}\footnote{We performed a 75-configuration grid search over \stableshots parameters and thresholds. We also empirically compared \stableshots with scaled Hoeffding and Weissman policies, showing that plain Hoeffding and Weissman bounds are too conservative, recommending millions of shots even for small circuits, while it is also impossible to find a good scaling factor (even for each specific algorithm--size pair) that reaches results as good as \stableshots. However, the strict 4-page limit does not leave enough space to discuss these experiments in detail. They will be presented in a future work. However, the full code, raw data, analysis scripts, and the frontier-exploration tool are freely available at: \url{https://github.com/GBisi/stableshots}} Our contributions are: (i) an auditable black-box shot-stopping rule for static circuit execution; (ii) a validation/test protocol for selecting and evaluating execution policies without using held-out traces during calibration; (iii) an empirical evaluation against fixed-shot baselines on QSimBench traces; and (iv) an open-source implementation and frontier-exploration tool.

\section{Related Work}
Analytical shot selection can be derived from concentration inequalities. Hoeffding's inequality bounds deviations of sums of bounded independent random variables~\cite{hoeffding1963probability}. For empirical distributions over finite alphabets, Weissman et al. bound the $L_1$ deviation between empirical and true distributions~\cite{weissman2003inequalities}, and later work refines related concentration results~\cite{mardia2018concentration}. Such bounds are valuable when formal guarantees are required, but their dependence on the output alphabet can make them conservative for full bitstring distributions. For $\tau=0.05$ and $\delta=0.05$, Hoeffding requires 82,707 shots at 4 qubits, 30.3 million at 8 qubits, and 179.8 billion at 14 qubits. Weissman is tighter but still reaches 2.27 million shots at 14 qubits. \stableshots instead provides an empirical online policy: it does not certify closeness to the unknown distribution, but it decides whether further batches still materially change the observed distribution.

Adaptive measurement allocation has also been studied for variational and iterative workloads. Methods for VQE and related algorithms distribute shots across Hamiltonian terms, gradients, observables, or optimizer steps~\cite{zhu2024optimizing,gu2021adaptive}; learning-based approaches can infer measurement-reduction policies from previous executions~\cite{liang2024artificial}. These techniques exploit algorithmic or observable structure. \stableshots is complementary: it targets a single static circuit and observes only streamed measurement counts. From a QSE perspective, this moves the shot budget from a pre-execution constant to an execution policy that can be calibrated, logged, and audited~\cite{bisicchia2025quantum,garcia2026rethinking}.

\section{StableShots}
Let $C$ be a fixed quantum circuit executed on backend $Q$. After $N$ shots, cumulative counts define an empirical distribution $\hat{P}_N$ over bitstrings. We compare empirical distributions using Total Variation Distance:
\begin{equation}
\tvd(\hat{P}_a,\hat{P}_b)=\frac{1}{2}\sum_x |\hat{P}_a(x)-\hat{P}_b(x)| .
\label{eq:tvd}
\end{equation}

\stableshots executes $C$ in batches of $b$ shots. After each batch, it compares $\hat{P}_N$ with $\hat{P}_{N-\ell b}$, where $\ell$ is the look-back in batches. If the resulting marginal TVD is at most $\epsilon$ for $k$ consecutive checks, execution stops; otherwise, execution continues until a maximum budget $B$ is reached. The parameters have direct operational meaning: $b$ controls decision granularity, $\ell$ controls the comparison window, and $(\epsilon,k)$ control how conservative the stability criterion is.

The rule is intentionally heuristic. It does not claim that $\hat{P}_N$ is within a certified distance from the unknown backend-induced distribution. Instead, it operationalizes a diminishing-returns signal: if several new batches have little effect on the cumulative empirical distribution, additional batches are unlikely to change the developer-visible result substantially. This makes the rule easy to implement and inspect. A run can be reconstructed from the batch sequence, the cumulative distributions, and the stopping predicate.

This design is deliberately conservative about what the wrapper assumes. It treats the circuit and backend as an opaque sampler and uses only the same information that a client already receives after execution: bitstring counts. Consequently, the method can be placed above different SDKs, simulators, or provider APIs without depending on compiler internals or hardware calibration schemas. The cost is that the stopping condition is local. It answers whether recent evidence has stabilized the cumulative empirical distribution, not whether a theorem certifies the final distribution. For the QSE use case addressed here, that distinction is useful: the policy is meant to support practical execution governance, where developers need an auditable stopping decision and an empirical record of the cost-fidelity trade-off.

For the experiments below we use the validation-selected profile \texttt{b50\_lb3\_k5\_eps0p005}: batches of 50 shots, a three-batch look-back, five consecutive stable checks, and threshold $\epsilon=0.005$. The external evaluation target is TVD $\leq0.05$ against a high-shot empirical reference. The configuration should be interpreted as dataset- and target-scoped, not universally optimal.

\section{Experimental Design}
We evaluate on 180 QSimBench traces~\cite{qsimbench}: six circuit families (\texttt{dj}, \texttt{qaoa}, \texttt{qft}, \texttt{qnn}, \texttt{random}, and \texttt{vqe}), six sizes from 4 to 14 qubits, and five noisy IBM simulated backends. Each trace contains 20,000 shots. The 20,000-shot empirical distribution is used only as an offline reference; \stableshots never observes it while deciding when to stop. We use the noisy-backend reference rather than an ideal simulator because the operational target of execution is the distribution induced by the selected backend, including noise~\cite{nielsen2010quantum}.

All strategies are evaluated on prefixes of the same materialized trace. This makes fixed budgets and adaptive stopping comparable on identical sample paths, but it also means that lower-shot estimates and the 20,000-shot reference are not independent. We therefore interpret TVD as a controlled empirical proxy rather than as an oracle error measure.

Configuration choices are made only on validation traces. We use 100 structured backend-holdout repetitions. In each repetition and for each algorithm-size cell, one backend is held out as test and the remaining four backends are used for validation. Since the benchmark contains six algorithms, six sizes, and five backends, each repetition has 144 validation traces and 36 test traces. After validation selects a configuration, we evaluate it on the held-out traces. We compare against fixed-shot baselines at 5,000, 10,000, and 18,000 shots, which summarize low, medium, and near-cap execution regimes from the original baseline set.

The holdout protocol is designed to match the operational question. A developer rarely knows in advance which backend-noise profile will make a circuit easy or hard to sample, but a tool builder can calibrate a policy on previous executions and then deploy it on future executions from the same workload family. Holding out one backend per algorithm-size cell tests this setting more strictly than a random split over traces: every test trace belongs to a circuit family and size seen during validation, but its backend-specific sample path was not used to choose the policy. We report success rates at TVD thresholds 0.01, 0.05, and 0.10, median and mean shots, maximum TVD, and how often the policy reaches the 20,000-shot cap. Reporting cap use is important because a policy can achieve high success simply by behaving like a near-reference fixed budget.

\section{Results}
Table~\ref{tab:overall} reports held-out test performance over 100 repetitions, for 3,600 test evaluations. Fixed-shot baselines improve as the budget grows, but no fixed budget provides the same cost-fidelity trade-off. Fixed-5k has the lowest cost but reaches TVD $\leq0.05$ on only 54.1\% of evaluations. Fixed-10k improves success to 66.7\%, while Fixed-18k reaches 83.3\% success but spends more than twice the median shots of \stableshots. The selected \stableshots profile reaches TVD $\leq0.05$ and $\leq0.10$ on all held-out evaluations with median 7,650 shots.

\begin{table}[t]
\centering
\caption{Held-out test results over 100 backend-holdout repetitions. TVD is measured against the 20,000-shot reference. The \stableshots row reports the selected configuration only; 20k cap reports maximum-budget use.}
\label{tab:overall}
\scriptsize
\setlength{\tabcolsep}{2.4pt}
\renewcommand{\arraystretch}{0.92}
\resizebox{\columnwidth}{!}{%
\begin{tabular}{lrrrrrrrr}
\toprule
Strategy & Med. shots & Mean shots & Med. TVD & Max TVD & $\leq$.01 & $\leq$.05 & $\leq$.10 & 20k cap \\
\midrule
Fixed-5k      & 5,000  & 5,000  & 0.0393 & 0.5608 & 14.1\% & 54.1\% & 67.2\% & 0.0\% \\
Fixed-10k     & 10,000 & 10,000 & 0.0236 & 0.3203 & 22.9\% & 66.7\% & 81.1\% & 0.0\% \\
Fixed-18k     & 18,000 & 18,000 & 0.0076 & 0.0891 & 55.9\% & 83.3\% & 100.0\% & 0.0\% \\
\stableshots  & 7,650  & 10,460 & 0.0193 & 0.0472 & 33.6\% & 100.0\% & 100.0\% & 27.7\% \\
\bottomrule
\end{tabular}%
}
\end{table}

The result should not be read as strict dominance. \stableshots can spend more shots than a low fixed budget because it continues when the empirical distribution is still moving. Conversely, it can accept larger TVD than near-reference fixed execution because it stops once the marginal distributional change is small. Its role is to provide an adaptive operating point between arbitrary low-shot execution and always using a high fixed budget.

Table~\ref{tab:groups} shows that the adaptive behavior is trace-dependent. Size and algorithm family dominate cap use. Small circuits usually stop early, while 10--14 qubit traces often continue toward the cap. The \texttt{dj} and \texttt{vqe} families rarely need the cap, whereas \texttt{qft}, \texttt{qnn}, and \texttt{random} often approach it. Backend variation is smaller: median shots range from 7,000 to 8,550, and every backend group remains below the 0.05 target. The same profile reaches TVD $\leq0.05$ on all 180 traces in the complete benchmark with median 7,700 shots, which is consistent with the held-out profile.

The cap-use pattern is a useful diagnostic rather than merely a cost statistic. When \stableshots reaches the cap, the policy is reporting that local stability was not obtained under the configured threshold and patience. This is preferable to silently returning a low-shot estimate whose empirical distribution is still changing. In the evaluated traces, cap use is concentrated in larger and more diffuse distributions, where fixed low budgets also show poor reliability. For users who tolerate coarser fidelity, the policy can be made more aggressive: in the original validation grid, \texttt{b50\_lb2\_k1\_eps0p005} reached TVD $\leq0.10$ on all traces with median 5,300 shots, but reached the stricter TVD $\leq0.05$ target on only 87.8\% of traces. This illustrates that \stableshots is not a single universal budget-reduction recipe; it is a tunable execution policy whose configuration should be tied to the target fidelity regime.

\begin{table*}[t]
\centering
\caption{Held-out structural profile of the selected configuration, aggregated over 100 backend-holdout repetitions. All entries use test traces only; all groups reach TVD $\leq0.05$. The last column reports use of the 20k cap.}
\label{tab:groups}
\scriptsize
\setlength{\tabcolsep}{2.6pt}
\renewcommand{\arraystretch}{0.90}
\begin{minipage}[t]{0.49\textwidth}
\centering
\resizebox{\linewidth}{!}{%
\begin{tabular}{llrrrrr}
\toprule
Factor & Group & Med. shots & Mean shots & Med. TVD & Max TVD & 20k cap \\
\midrule
Size & 4 & 2,750 & 2,684 & 0.0150 & 0.0272 & 0.0\% \\
Size & 6 & 6,725 & 5,536 & 0.0230 & 0.0315 & 0.0\% \\
Size & 8 & 11,500 & 10,529 & 0.0230 & 0.0338 & 0.0\% \\
Size & 10 & 19,075 & 14,025 & 0.0094 & 0.0346 & 50.0\% \\
Size & 12 & 20,000 & 14,703 & 0.0000 & 0.0357 & 56.2\% \\
Size & 14 & 20,000 & 15,284 & 0.0000 & 0.0472 & 59.8\% \\
\bottomrule
\end{tabular}%
}
\end{minipage}
\hfill
\begin{minipage}[t]{0.49\textwidth}
\centering
\resizebox{\linewidth}{!}{%
\begin{tabular}{llrrrrr}
\toprule
Factor & Group & Med. shots & Mean shots & Med. TVD & Max TVD & 20k cap \\
\midrule
Alg. & \texttt{dj} & 3,500 & 3,389 & 0.0228 & 0.0418 & 0.0\% \\
Alg. & \texttt{qaoa} & 12,725 & 12,053 & 0.0249 & 0.0346 & 16.0\% \\
Alg. & \texttt{qft} & 18,725 & 14,954 & 0.0088 & 0.0308 & 50.0\% \\
Alg. & \texttt{qnn} & 17,750 & 14,295 & 0.0079 & 0.0280 & 50.0\% \\
Alg. & \texttt{random} & 18,450 & 14,319 & 0.0049 & 0.0298 & 50.0\% \\
Alg. & \texttt{vqe} & 3,750 & 3,754 & 0.0233 & 0.0472 & 0.0\% \\
\bottomrule
\end{tabular}%
}
\end{minipage}

\vspace{0.20em}
\resizebox{0.70\textwidth}{!}{%
\begin{tabular}{llrrrrr}
\toprule
Factor & Group & Med. shots & Mean shots & Med. TVD & Max TVD & 20k cap \\
\midrule
Backend & \texttt{fez} & 7,000 & 10,102 & 0.0225 & 0.0430 & 24.3\% \\
Backend & \texttt{kyiv} & 8,550 & 11,232 & 0.0158 & 0.0454 & 32.8\% \\
Backend & \texttt{marrakesh} & 7,350 & 9,337 & 0.0218 & 0.0407 & 22.6\% \\
Backend & \texttt{sherbrooke} & 8,050 & 10,823 & 0.0186 & 0.0472 & 29.1\% \\
Backend & \texttt{torino} & 7,950 & 10,785 & 0.0188 & 0.0427 & 29.4\% \\
\bottomrule
\end{tabular}%
}
\vspace{-0.8em}
\end{table*}

These data clarify what the stopping signal captures. It is not merely a proxy for qubit count: circuits of the same size can stop at different budgets, and backend effects are visible but secondary. The policy therefore exposes a useful execution-level signal for quantum software tooling. Instead of asking developers to choose one budget for all workloads, a runtime or client library can record batch counts, apply a calibrated stopping rule, and report both the final distribution and the evidence that triggered termination.

\section{Discussion and Limitations}
The objective of \stableshots is not to replace statistical guarantees with a universal empirical claim. The contribution is an execution policy and an evaluation workflow: the policy defines a transparent online signal, and the validation/test split prevents the reported configuration from being justified only on the traces used to choose it. This framing fits QSE practice because shot count becomes part of the software execution process: it can be parameterized, calibrated, logged, and revised as workloads or backends change.

\textbf{Practical applicability.} \stableshots assumes that a client can obtain batches of counts from repeated executions of the same circuit. On some hardware services, submitting many tiny independent jobs would be inefficient because queueing and job-management overhead could dominate measurement savings. However, current quantum-cloud execution models increasingly support iterative workloads. IBM Runtime\footnote{ https://quantum.cloud.ibm.com/docs/en/guides/run-jobs-session} sessions group iterative calls and prioritize subsequent jobs within the session, while caching data that can reduce repeated overhead. Amazon Braket reservations\footnote{https://docs.aws.amazon.com/braket/latest/developerguide/braket-reservations.html} provide exclusive access to a selected device for a scheduled time window, making it possible to run multiple tasks without repeatedly competing in the global queue. These mechanisms do not eliminate the need to choose a practical batch size, but they make incremental execution plausible when batches are submitted inside a session, reservation, or provider-supported batch workflow.

\textbf{Limitations.} First, TVD is measured against a finite 20,000-shot reference, not the unknown backend-induced distribution. This is appropriate for noisy backends, where the noiseless ideal distribution is not the operational target, but the reference is still finite. Second, fixed budgets, adaptive prefixes, and references are derived from the same materialized traces. This improves comparability across policies but couples estimates and may favor longer prefixes. Third, the experiments use noisy simulated QSimBench backends; live QPU behavior, queueing, and backend drift require separate evaluation. Fourth, we assume that the backend-induced distribution is approximately stable during the execution of one circuit trace, so online stopping can be interpreted as convergence toward a fixed but unknown target distribution. This assumption is consistent with prior empirical evidence~\cite{bisicchia2025quantum}, but stronger temporal drift on live hardware could invalidate it. Fifth, \stableshots targets static circuits. Within a variational iteration the parameters are fixed and \stableshots could be used locally, but non-stationarity across optimizer steps would require additional mechanisms.

These limitations define the boundary of the present claim. The result is not that \stableshots provides a distribution-free guarantee, nor that the reported parameter tuple should be used unchanged in all applications. The result is that a simple black-box stability signal can be selected without test leakage and can outperform common fixed-shot choices under a practical cost-fidelity objective. That is the core QSE contribution: execution cost is treated as a controllable software policy, and the policy is evaluated with the same empirical discipline expected for other quantum software tools.

A practical adoption path is to expose \stableshots as an execution mode rather than as a replacement for fixed shots. A developer could request a target policy, maximum cap, and batch size; the runtime would then return the final counts, the consumed shots, the termination reason, and the stability trace. This keeps the familiar fixed-shot interface available for reproducibility-critical experiments, while adding an adaptive option for exploratory execution, benchmarking, regression testing, and service-oriented quantum workflows. It also makes failures informative: if a circuit repeatedly reaches the cap, the tool has evidence that the chosen target is too strict for the allowed budget or that the circuit-backend pair produces a distribution that stabilizes slowly.

This reporting discipline is important for empirical QSE. Shot count is often treated as an incidental parameter in experimental papers and tool evaluations, even though it can change both cost and measured quality. By making the stopping decision explicit, \stableshots encourages authors and tool builders to report not only final accuracy but also how the measurement budget was consumed. In the long term, such execution metadata can support benchmark repositories in which shot policies are compared across workloads, reused under version control, and recalibrated when new backends or provider execution models become available.

\section{Conclusions and Future Work}
We presented \stableshots, an online shot-stopping rule for static quantum circuit execution. The method monitors changes in cumulative empirical output distributions and stops when additional batches have repeatedly small marginal effect. On 180 QSimBench traces and 100 backend-holdout repetitions, the selected configuration reaches TVD $\leq0.05$ on all held-out test evaluations with median 7,650 shots, while fixed-shot baselines either under-execute more often or spend substantially more measurements.

Future work should evaluate live QPU executions, independent high-shot references, and adaptive batch sizes. Richer policies could combine distributional change with support growth, uncertainty estimates, or backend-drift signals. For iterative workloads such as VQA and QML, \stableshots could be used locally within each fixed-parameter evaluation while exploiting warm-start information across iterations. Another direction is to combine online stopping with circuit cutting and quantum multiprogramming, maintaining independent stopping states for subcircuits or co-scheduled circuits so that shots are spent where empirical distributions have not yet stabilized.


\begin{thebibliography}{15}

\bibitem{nielsen2010quantum}
M.~A. Nielsen and I.~L. Chuang, \emph{Quantum Computation and Quantum Information}. Cambridge University Press, 2010.

\bibitem{bisicchia2024quantum}
G.~Bisicchia et al., ``From quantum software handcrafting to quantum software engineering,'' in \emph{Proc. IEEE SANER-C}, 2024.

\bibitem{hoeffding1963probability}
W.~Hoeffding, ``Probability inequalities for sums of bounded random variables,'' \emph{Journal of the American Statistical Association}, 1963.

\bibitem{weissman2003inequalities}
T.~Weissman et al., ``Inequalities for the $L_1$ deviation of the empirical distribution,'' Hewlett-Packard Labs, Tech. Rep., 2003.

\bibitem{mardia2018concentration}
J.~Mardia et al., ``Concentration inequalities for the empirical distribution,'' arXiv:1809.06522, 2018.

\bibitem{zhu2024optimizing}
L.~Zhu et al., ``Optimizing shot assignment in variational quantum eigensolver measurement,'' \emph{Journal of Chemical Theory and Computation}, 2024.

\bibitem{gu2021adaptive}
A.~Gu et al., ``Adaptive shot allocation for fast convergence in variational quantum algorithms,'' arXiv:2108.10434, 2021.

\bibitem{liang2024artificial}
S.~Liang et al., ``Artificial-intelligence-driven shot reduction in quantum measurement,'' \emph{Chemical Physics Reviews}, 2024.

\bibitem{bisicchia2025quantum}
G.~Bisicchia et al., ``Distributing quantum computations, shot-wise,'' in \emph{Future Internet}, 2025.

\bibitem{garcia2026rethinking}
J.~Garcia-Alonso et al., ``Rethinking Services in the Quantum Age: The SOQ Paradigm,'' in \emph{ACM Transactions on Software Engineering and Methodology}, 2026.

\bibitem{qsimbench}
G.~Bisicchia et al., ``QSimBench: An execution-level benchmark suite for quantum software engineering,'' in \emph{Proc. IEEE QCE}, 2025.

\end{thebibliography}
\end{document}